\documentclass{article}
\usepackage{tikz}
\usepackage{amsthm,amsmath,amssymb,array,color,colortbl,graphicx,multirow}
\usepackage{comment}
\usepackage{algorithm}
\usepackage{balance}
\usepackage{xspace}
\usepackage{wrapfig}
\usepackage{enumitem}
\usepackage{fullpage}

\definecolor{Gray}{gray}{0}
\definecolor{RED}{RGB}{255,0,0}
\definecolor{GREEN}{RGB}{0,255,0}
\definecolor{BLUE}{RGB}{0,0,255}
\definecolor{YELLOW}{RGB}{255,255,0}
\definecolor{BLACK}{RGB}{0,0,0}

\newtheorem{theorem}{Theorem}

\def\comp{\Psi}
\def\row{{\Gamma}}

\def\uni{\mathcal{U}}

\usepackage[english]{babel}
\usepackage{color}
\usepackage{url}
\usepackage{subfig}
\usepackage{xspace}
\usepackage{amssymb} 
\usepackage{amsmath} 
\usepackage{amsthm}
\usepackage{float}
\usepackage{graphicx}
\usepackage{footnote}
\usepackage{comment}
\makesavenoteenv{tabular}
\makesavenoteenv{table}
\usepackage{multirow}
\usepackage{tikz}
\usetikzlibrary{calc}
\usepackage{algpseudocode}
\usepackage[acronym]{glossaries}

\newcommand{\para}[1]{\noindent \textbf{#1}\xspace}

\begin{document}

\title{Measuring the Complexity of Packet Traces\thanks{Authors appear in alphabetical order.
Research conducted as part of Chen Griner's thesis.}}

\author{Chen Avin$^1$ \quad Manya Ghobadi$^2$ \quad Chen Griner$^1$ \quad Stefan Schmid$^3$ \\
\small $^1$ School of Electrical and Computer Engineering, Ben Gurion University of the Negev, Israel \\
\small $^2$ Computer Science and Artificial Intelligence Laboratory, MIT, USA \\
\small $^3$ Faculty of Computer Science, University of Vienna, Austria
}
\date{}
\maketitle

\begin{abstract}
This paper studies the structure of several real-world traces (including Facebook,  High Performance Computing, Machine Learning, and simulation generated traces) and presents a systematic approach to quantify and compare the structure of packet traces based on the entropy contained in the trace file. Insights into the structure of packet traces can lead to improved network algorithms that are optimized
toward specific traffic patterns. We then present a  methodology  to quantify the temporal and  non-temporal components of entropy contained in a packet trace, called the \emph{trace complexity}, using randomization and compression. We show that trace complexity provides unique insights into the characteristics of various applications and argue that there is a need for traffic generation models that preserve the intrinsic structure of empirically measured application traces. We then propose a traffic generator model that is able to produce a synthetic trace that matches the complexity level of its corresponding real-world trace. 
\end{abstract}

\section{Introduction}\label{sec:intro}

Packet traces collected from networking applications, such as data center traffic, tend not to be \emph{completely random} and have been observed to feature \emph{structure}:
data center traffic matrices are sparse and skewed~\cite{benson2010network, projector}, 
exhibit locality~\cite{chen2014osa}, and are bursty~\cite{DBLP:journals/cn/ZouW0HCLXH14, datacenter_burstiness}.
Motivated by the existence of
such structure, the networking community is currently 
putting much effort into designing algorithms to optimize different network layers toward such structure towards self-driving and demand-aware networks~\cite{ccr18san, deepq, group_tracking}, learning-based traffic engineering~\cite{learning_to_route} and video streaming~\cite{pensieve}, as well as reconfigurable optical
networks~\cite{rotornet, projector, firefly}. 
For instance, many 
network optimizations exploit the presence
of elephant flows~\cite{vl2, alizadeh2013pfabric}.

However, the structure available in different applications can differ significantly, and a unified approach to measure the structure in traffic traces is missing. Better quantification of a trace structure will lead to better network optimization and to the understanding of the available improvement, if possible, in current solutions. Moreover, one of the critical factors in evaluating new proposals is their traffic workload. Ideally, the traffic workload should contain the same structure as real-world traces but often is overlooked due to the lack of a traffic generation model that can replicate real-world traces while preserving their temporal and non-temporal structure.

\begin{figure}
  \begin{centering}
    \includegraphics[width=.8\columnwidth]{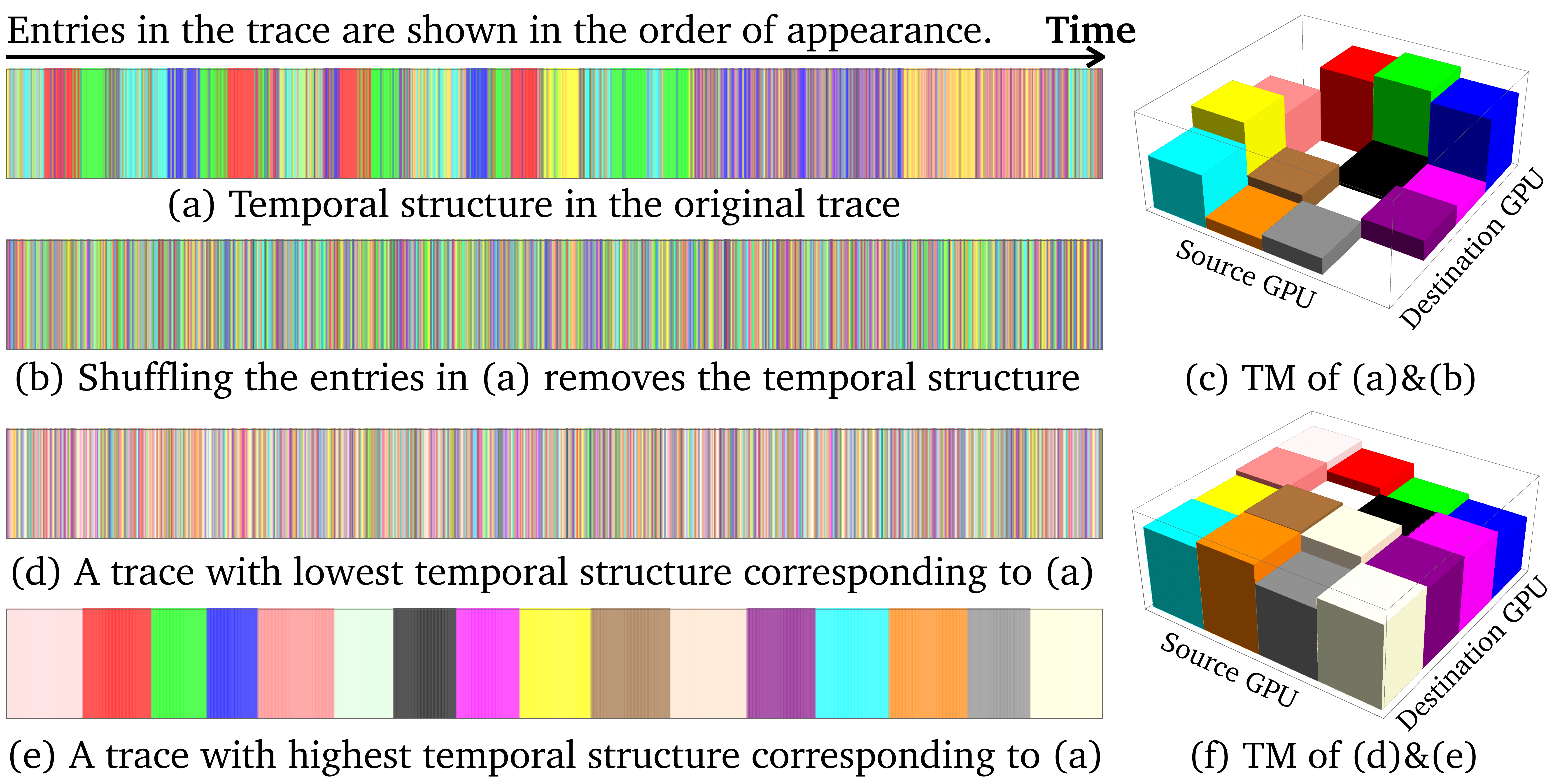}
    \caption{Visualization of temporal and non-temporal structure in machine learning workload.}
    \label{fig:intro2}
  \end{centering}
\end{figure}

For instance, consider a trace file including the communication pattern of a Machine Learning (ML) application executing a popular convolutional neural network training job on four GPUs. This workload was obtained from authors of ~\cite{sip-ml}. 
Figure~\ref{fig:intro2}(a) is a visualization of the trace file where each entry in the trace is represented by a unique color corresponding to its <source, destination> GPU pair. This visualization shows the \emph{temporal} structure in the trace file, as colors appear consecutive and follow some pattern. In contrast, Fig.~\ref{fig:intro2}(b) shows the same trace file but the entries in the file are shuffled to remove the temporal structure in Fig.~\ref{fig:intro2}(a). The traffic matrix (TM) in Fig.~\ref{fig:intro2}(c) shows a skewed heatmap indicating that some GPU pairs communicate more frequently than others. Note that even thought the temporal structures in (a) and (b) are different, they both have the same TM shown in (c). In other words, the TM is able to capture the non-temporal structure in the trace files but not the temporal one. For comparison, let us now consider two synthetic traces shown in Fig.~\ref{fig:intro2}(d) and (e). Trace (d) is generated uniformly and random and has the lowest temporal structure compared to (a), while trace (e) is sorted based on <source, destination> key and hence has the highest temporal structure. Similarly, Fig.~\ref{fig:intro2}(f) captures the non-temporal structure in (d) and (e) but not the temporal one.

But how can we \emph{measure} the structure of a trace file? And how should we generate synthetic traces while preserving the structure of empirical traces? This paper aims at providing initial steps to these questions. In particular, we quantify the amount of temporal and non-temporal structure in traffic traces using the information theoretic measure of \emph{entropy}~\cite{shannon1948mathematical} in the trace. Since the term entropy is define for random variables, as opposed to a sequence of individual communication requests in a packet trace, we use a more general term called  complexity~\cite{ziv1978compression} to quantify the structure in a packet trace and call it \emph{trace complexity}. Moreover, we provide a traffic generation model to generate synthetic traces that match the structure of a given trace.
Intuitively, a packet trace with \emph{high structure} has \emph{low entropy} and \emph{low complexity}: it contains little information, and the sequence behavior is more predictable \cite{feder1992universal}.  Our approach allows us to chart, what we call, a \emph{complexity map} of individual traffic traces: to map each traffic trace to a two-dimensional graph indicating the amount of temporal and non-temporal information that is present in a trace.  

The main contributions of this paper are as follows. First, we present an information theoretic perspective to systematically separate the temporal and non-temporal structures available in a traffic trace (\S\ref{sec:methodology}). Second, we compare the complexity of 17 trace files, shown in Table~\ref{tab:traces_data}, including production-level traces (Facebook~\cite{roy2015inside} and High Performance Computing~\cite{doe2016characterization}), NS2 simulation-based traces (pFabric~\cite{alizadeh2013pfabric}), Machine Learning workload (SiP-ML~\cite{sip-ml}), as well as our own generated traces for reference points (for uniform, skewed, and bursty traffic patterns). Our results indicate that production-level traces have high temporal complexity while NS2 generated traces have high non-temporal structure. Third, we show that our methodology can be used to slice the complexity into rack-level versus IP-level, as well as source versus destination structures (\S\ref{sec:map-takeaways}). Finally, we present a simple yet powerful model to generate traffic traces that match the complexity of production-level traces (\S\ref{sec:model}).

\section{Defining Trace Complexity}
\label{sec:methodology}

This section describes our methodology to quantify the inherent structure in packet traces. Given a packet trace, $\sigma$, we define its \emph{trace complexity} as the ratio of its entropy over that of a random trace, $\uni(\sigma)$. Intuitively, a random trace does not compress as well as a more structured trace. At the heart of our methodology lie two main concepts: ($i$) \emph{Randomization:} we systematically randomize different slices of a traffic trace to profile their contributions to the trace complexity; ($ii$) \emph{Compression:} we then measure the complexity of the trace and its randomized variants by compressing the trace files. The size of the compressed trace file is then taken as the measure of trace complexity. Next, we explain these two concepts more formally.

\para{Eliminate Structure by Randomization.} A traffic trace file $\sigma$ consists of an ordered list of entries $\sigma_1, \sigma_2, \dots, \sigma_t$, where each entry $\sigma_i = (s_i, d_i)$ is a <source, destination> pair arriving at time $i$. In this work, we ignore other fields in a packet header (such as packet size and port number) and focus on the order of entries in the trace file, to capture temporal complexity, and source/destination pairs, to capture non-temporal complexity, of the trace. We find that there is enough information in our methodology to capture the differences in temporal and non-temporal complexities of real traces. For instance, \S\ref{sec:map} shows that Facebook's Hadoop trace~\cite{roy2015inside} has about 50\% less temporal structure compared to pFabric's trace~\cite{alizadeh2013pfabric}.

\para{Trace Complexity.} We now define the trace complexity, $\comp(\sigma)$, as the ratio between the complexity of the original trace, $\sigma$, to the expected complexity of its randomized counterpart, $\uni(\sigma)$, where each of its entries are chosen \emph{uniformly at random} from the set of IDs in $\sigma$:
\begin{equation}
\comp(\sigma) = \frac{C(\sigma)}{C(\uni(\sigma))}.
\label{eq:trace_complexity}
\end{equation}
As we will discuss later in this section, $C(\cdot)$ represents the size of the compressed trace file and the more structure a trace has, the better it can be compressed. Hence, $\comp(\sigma) \in [0,1]$ since $C(\sigma) \le C(\uni(\sigma))$.

\begin{table}[t]
\caption{Traces used in the paper.}
\vspace{-0.4cm}
    \centering
    \begin{tabular}{|l|c|c|c|c|}
        \hline
        Type & \# traces & Sources & Dests & Avg.\# of entries  \\ \hline 
         Machine Learning~\cite{sip-ml}           & 1     & 4         &  4    & 1869    \\ \hline
         Facebook (IP) \cite{roy2015inside} & 3     & 174K       & 156K & 31M   \\ \hline  
         Facebook (Rack) \cite{roy2015inside}& 3     & 282       & 15K & 31M   \\ \hline  
         HPC  \cite{doe2016characterization}& 3     & 1024      & 1024  & 11M   \\ \hline  
         pFabric \cite{alizadeh2013pfabric} & 3     & 144       & 144 & 30M   \\ \hline  
         Reference Points (own)             & 4     & 16        & 16    & 10M   \\ \hline  
    \end{tabular}
    \label{tab:traces_data}
\end{table}

\para{Temporal Trace Complexity.} The temporal structure of a trace is a reflection of the burstiness in the traffic pattern. Prior work has measured the degree of burstiness in a trace as a sequence data packets with inter-arrival time less or
equal to 1 millisecond~\cite{source-level-burstiness, trickle, datacenter_burstiness}. Instead, we capture the temporal structure in a trace, $\sigma$, by systematically randomizing the original trace to eliminate all temporal relations in the trace and obtain a new trace file $\row(\sigma)$. Intuitively, $\row(\sigma)$ is a trace where each communication request is chosen independently at random from the previous ones. Formally, let  $\row(\sigma)$ be a temporal transformation: a transformation that performs a uniform random permutation of the rows of $\sigma$, eliminating any time dependency between rows, therefore $C(\sigma) \le C(\row(\sigma))$.
To measure how  much temporal complexity is contained in $\sigma$, we therefore normalized it by the complexity of its temporal transformation, $\row(\sigma)$. Hence, the normalized temporal trace complexity is defined as $T(\sigma) = \frac{C(\sigma)}{C(\row(\sigma))} \in [0,1]$.

\begin{figure}[t]
  \begin{centering}
    \includegraphics[width=0.7\columnwidth]{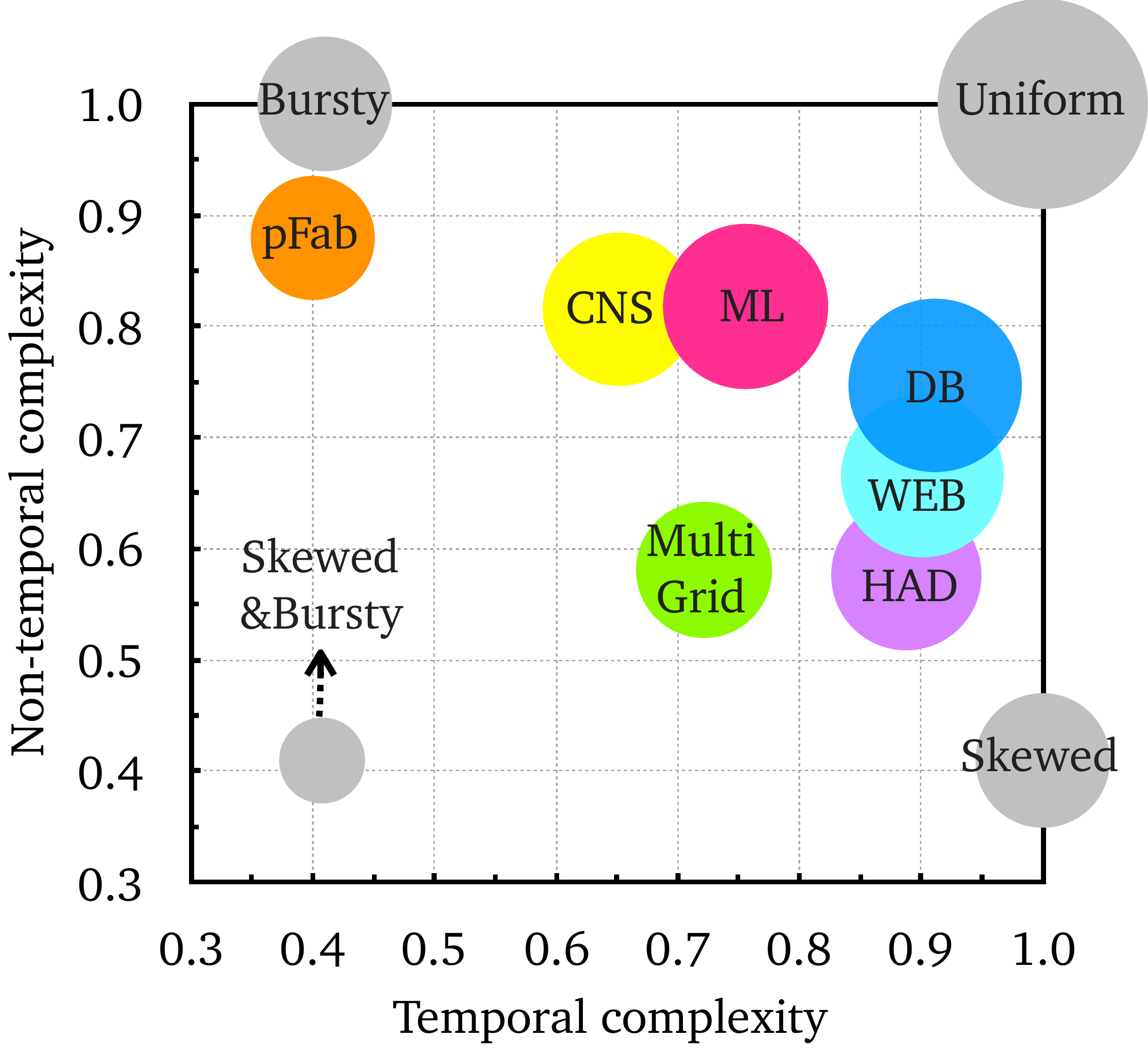}
    \caption{The complexity map of seven real traces (colored circles) and four reference points placed on the corners of the map.}
    \label{fig:example_complexity_map}
  \end{centering}
\end{figure}

\para{Non-Temporal Trace Complexity.} Note that, non-temporal structure is unaffected by the transformation $\row(\sigma)$ because correlations such as requests frequency, source destination dependency are conserved, while only the \emph{order} of the elements is changed such that it is uniformly random. Therefore, all the remaining complexity after the elimination of temporal complexity is non-temporal. This can be formally defined using our methodology by normalizing $\row(\sigma)$ with $\uni(\sigma)$ which has maximum complexity and no structure: $NT(\sigma) = \frac{C(\row(\sigma)))}{C(\uni(\sigma))} \in [0,1]$.

\para{Theoretical Properties.} It directly follows from our definition that the measure of trace complexity $\comp(\sigma)$ defined in Eq.~\ref{eq:trace_complexity} is the multiplication of the temporal and non-temporal complexity ratios. Formally, $\comp(\sigma) = T(\sigma) \times NT(\sigma)$. An important feature of our methodology is that is 
enables comparing traces of different sizes and domains. Section \ref{sec:formal_guarantee} describes the relationship of our metric to the entropy rate of a trace when $\sigma$ is generated by a stationary stochastic process.

\para{Compression-based Complexity.} Our methodology to measure the empirical entropy of a trace file relies on the principle of \emph{data compression}. The better we can compress a traffic trace, $\sigma$, the lower must be its entropy rate and hence its complexity. We assume a compression function or algorithm $C$ is applied to $\sigma$ and the complexity of $\sigma$, $C(\sigma)$, is the size of the compressed trace file. In this work, we use the 7zip compressor with Lempel-Ziv-Markov chain compression (LZMA)~\cite{7zip}; other compression techniques such as DEFLATE~\cite{deutsch1996deflate} can also be used.

\section{Quantifying the Complexity of Real-world Traces}
\label{sec:map-takeaways}

In this section, we first propose a graphical representation, called the \emph{complexity map}, to quantify and compare the temporal and non-temporal complexities of different traces on a 2-dimensional plane (\S\ref{sec:map}). We then use the complexity map to draw three main takeaways of our analysis on real-world traces (\S\ref{subsec:takeaways}).

\para{Setup and Dataset.} As shown in Table~\ref{tab:traces_data}, our dataset consists of 17 trace files in five categories: ($i$) trace files from three Facebook (FB)~\cite{roy2015inside} datacenters: hadoop (HAD), web (WEB), and database (DB) including IP and rack-level traces; ($ii$) MPI traces of three exascale applications in high performance computing (HPC) clusters~\cite{doe2016characterization}: CNS, MultiGrid, and NeckBone; ($iii$) pFabric~\cite{alizadeh2013pfabric} packet traces that we generated by running the NS2 simulation script obtained from the authors of the paper; ($iv$) a machine learning (ML) trace we obtained from~\cite{sip-ml} that measures the communication pattern between four GPUs running VGG19, a popular convolutional neural network training job; and $(v)$ four reference traces that we synthetically generate to represent bursty, skewed, busty \& skewed, and uniform traces. To avoid result distortion due to non-random ID selection in production traces, we uniformly hash all the source/destination IDs to the same length and domain.

\begin{figure}[t]
 \begin{centering}
  \begin{tabular}[t]{ccc}
  \includegraphics[width=.25\columnwidth]{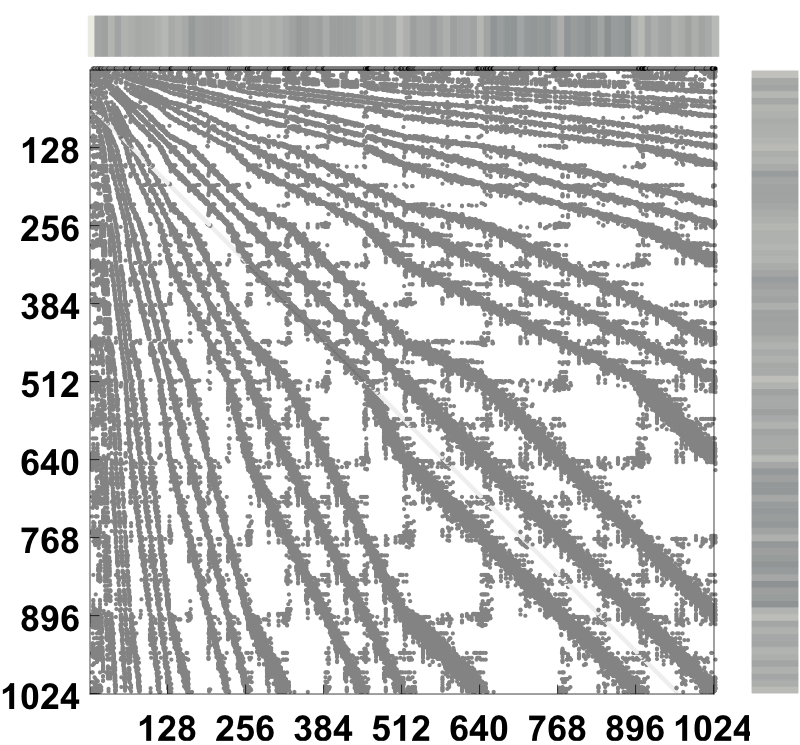} &
   \includegraphics[width=.25\columnwidth]{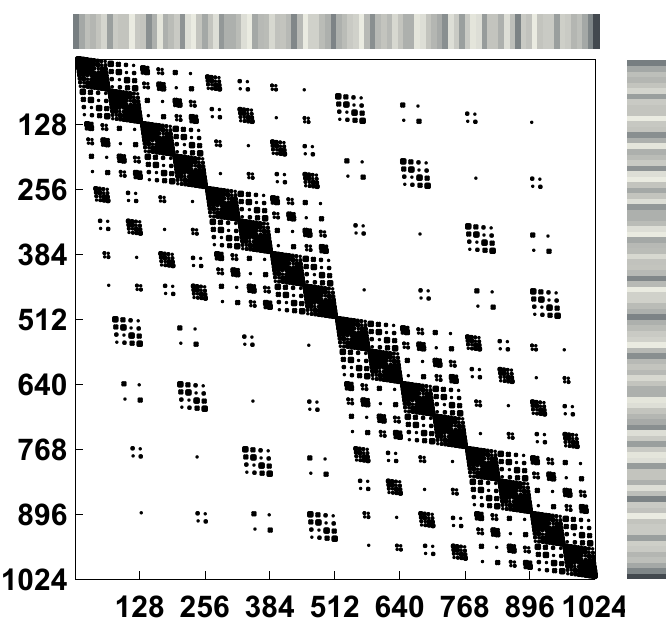} &
   \includegraphics[width=.25\columnwidth]{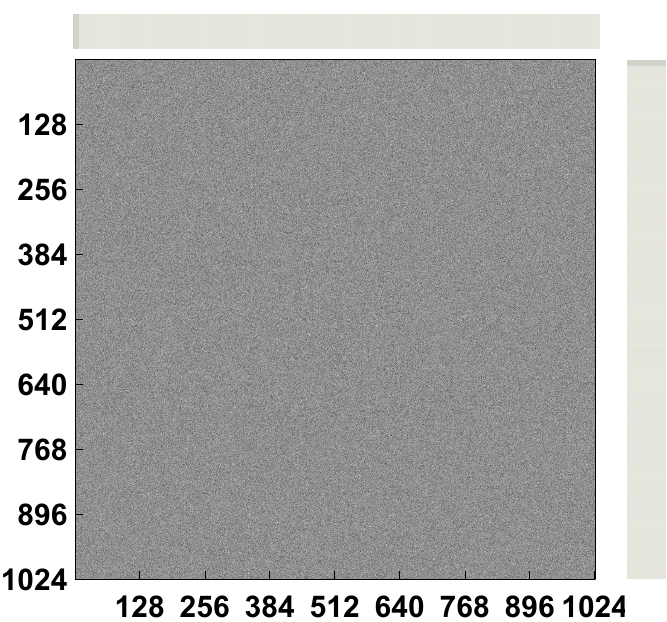} \\
    (a) & (b) & (c)
\end{tabular}
    \caption{Traffic matrices corresponding to three traces in Fig.~\ref{fig:example_complexity_map}: (a) CNS application, (b) MultiGrid application, (c) Uniform reference point.}
    \label{fig:HPCmat}
 \end{centering}
\end{figure}

\subsection{The Complexity Map} 
\label{sec:map}
To compare the complexity of traces in our dataset, we place them on a complexity map where the X and Y axes represent the temporal and non-temporal complexity dimensions. Fig.~\ref{fig:example_complexity_map} shows the complexity map of seven real traces and four reference points where each circle indicates a trace and the area of the circles corresponds to the overall complexity of $\sigma$, $\comp(\sigma)$, which is calculated by multiplying the temporal and  non-temporal complexities of~$\sigma$, as described in \S\ref{sec:methodology}.

Before diving into our main takeaways, we describe the four reference points on the complexity map. 
These points indicate the theoretical complexity of four hypothetical synthetic traces labeled as \emph{Uniform}, \emph{Skewed}, \emph{Bursty} , and \emph{Skewed \& Bursty}. The \emph{Uniform} trace is located at (1,1), indicating that it has the highest possible complexity and, hence, no structure. This means that the trace is a uniformly chosen random sequence. The \emph{Skewed} trace is located at (1,0.4), which indicates that it has high temporal complexity and low non-temporal complexity. This is a result of requests in the sequence that are distributed \emph{iid} (and hence with high temporal complexity), but which arrive from a skewed distribution with low entropy (and hence have low non-temporal complexity). In contrast, the \emph{Bursty} trace, located at (0.4,1), has low temporal complexity and high non-temporal complexity. This is the case when the next source-destination pair is selected uniformly at random, i.e., with high non-temporal complexity, but then repeated for some time (i.e., modeling a burst), creating temporal patterns and lower temporal complexity. Lastly, the \emph{Skewed \& Bursty} trace, located at (0.4, 0.4), has the lowest complexity in the current map and has both temporal and non-temporal structure. Requests are both temporally dependent (i.e., with repetitions) and new requests arrive from a skewed distribution. All four traces can be generated using a Markovian model which we describe in more details later in Section~\ref{sec:model}. Fig.~\ref{fig:HPCmat} shows the traffic matrix of three of the traces in Fig.~\ref{fig:example_complexity_map}: CNS, MultiGrid, and the uniform reference point. The observation that MultiGrid has less non-temporal complexity than CNS is captured by the differences in their traffic matrices shown in Fig.~\ref{fig:HPCmat}. In particular, we can observe that Fig.~\ref{fig:HPCmat}(a) has less structure than \ref{fig:HPCmat}(b), hence CNS has higher non-temporal complexity than MultiGrid. In contrast, Fig.~\ref{fig:HPCmat}(c) shows no structure and hence it has the highest complexity in the complexity map in Fig~\ref{fig:example_complexity_map}.

\vspace{-0.2cm}
\subsection{Takeaways}
\label{subsec:takeaways}

In this section, we apply the complexity map to different traces and
discuss the main takeaways with respect to their complexities and the differences between them.

\begin{figure}
  \begin{centering}
   \begin{tabular}{ccc}
    \includegraphics[width=.3\textwidth]{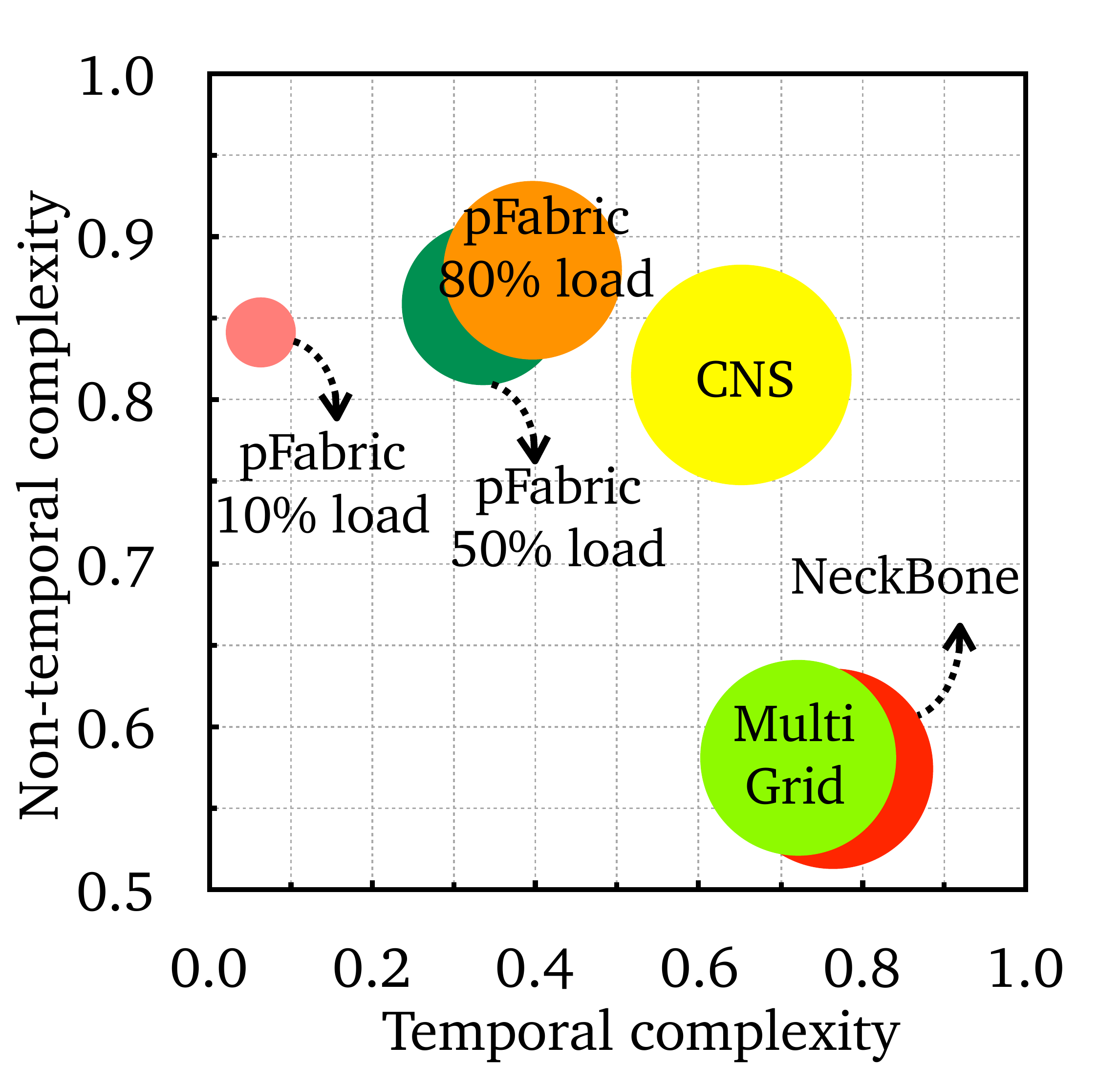} &
    \includegraphics[width=.3\textwidth]{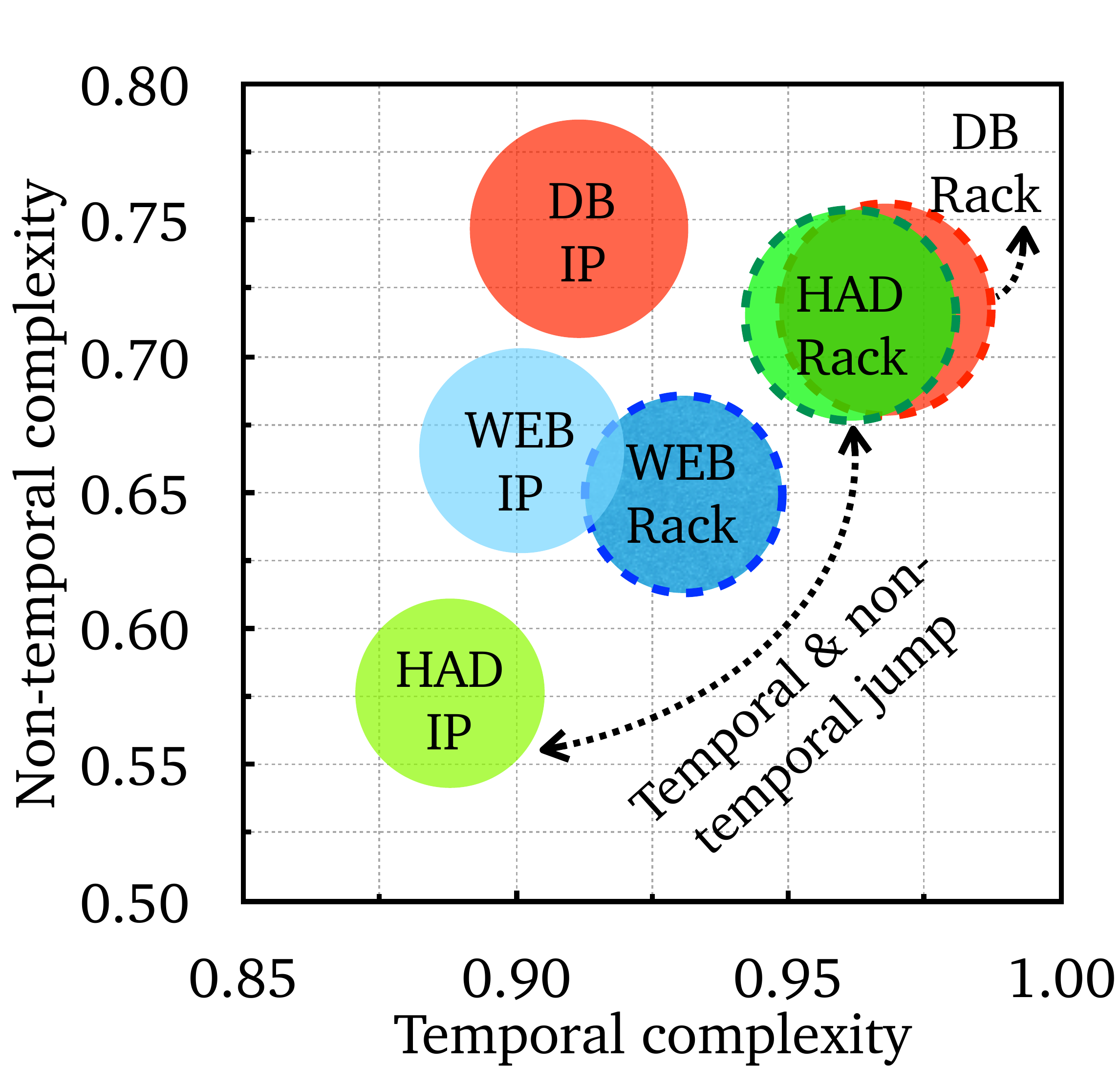} &
    \includegraphics[width=.3\textwidth]{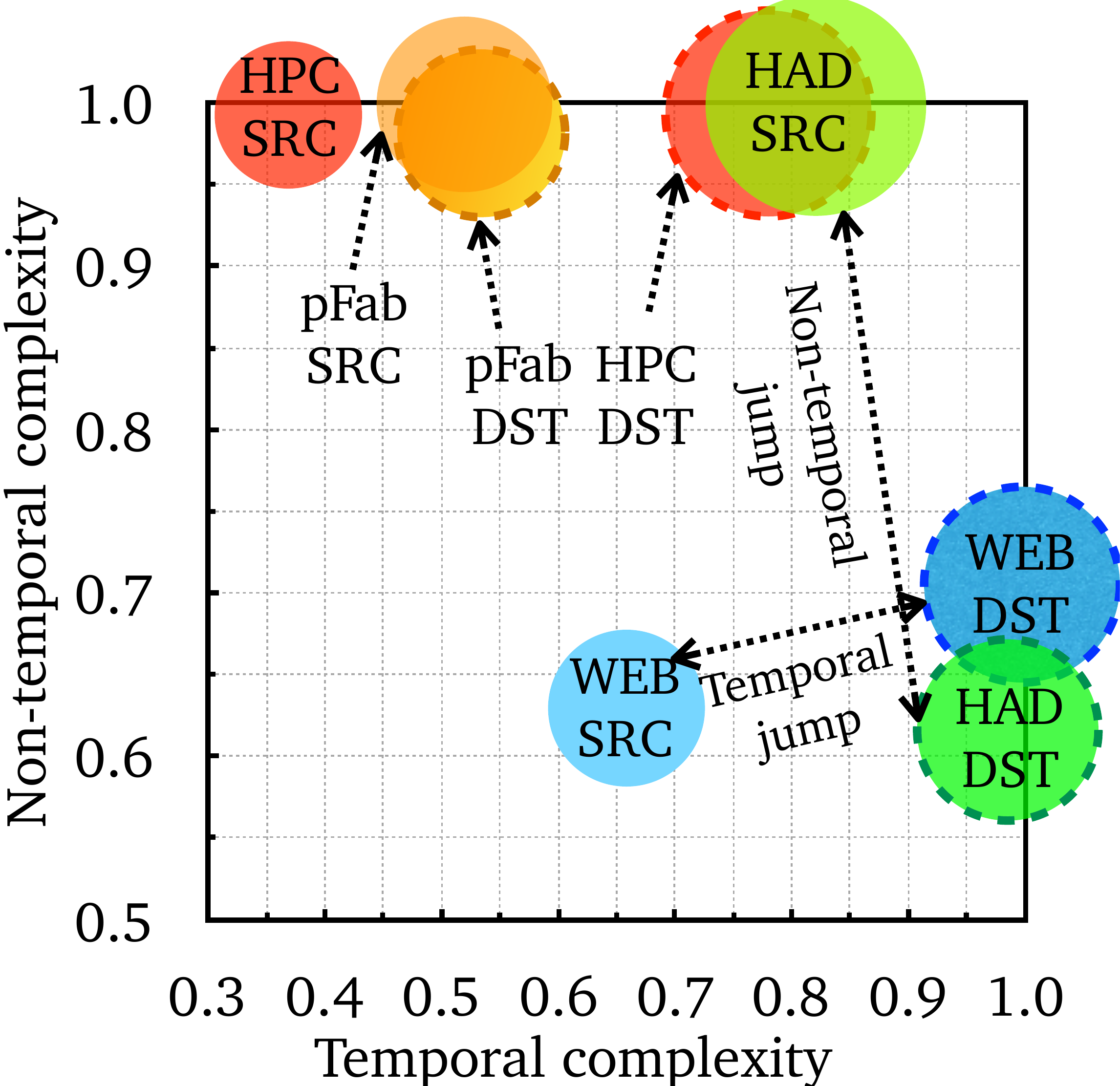} \\ 
    (a) pFabric \& HPC traces & (b) IP and Rack-level aggregations & (c) Src\&Dst-level complexities
   \end{tabular}
    \caption{Using different complexity maps to understand various trace structures.}
    \label{fig:takeaways}
  \end{centering}
\end{figure}

\para{Applications have different complexity measures.}
The complexity map highlights the different characteristics
and structures available in different applications, confirming
observations such as~\cite{roy2015inside} conducted on Facebook's datacenters.
Recall Fig.~\ref{fig:example_complexity_map}: pFabric and ML
traces feature a higher non-temporal complexity than MultiGrid and all Facebook (DB, HAD, WEB) traces, but pFabric has a lower temporal complexity than all the other traces.
Interestingly, Facebook traces have the highest temporal complexity. We suspect this is because of the 30,000 to 1 sampling of Facebook traces that destroys the temporal structure resulting in a high temporal complexity. This indicates that different applications may be identified by their specific 
complexity characteristic, and may provide different
opportunities for optimization.
To obtain a more detailed understanding, let us zoom in to the
pFabric and HPC traces. Fig.~\ref{fig:takeaways}(a) shows
that the complexity of pFabric also depends on the load (here, 10\%, 50\%, and 80\% loads are shown): at lower loads, fewer flows are competing, and hence flows mix to a lesser extent, naturally resulting in a lower temporal complexity. 
While this is expected, it validates our methodology and shows that compression can capture this behavior.
The non-temporal complexity of different HPC traces (CNS, MultiGrid, and NeckBone) depends on the specific application,
but is generally lower than that of pFabric, validating that compression can capture the non-temporal structure, 
as expected.

\para{Aggregation-level matters.}
The Facebook traces provide an opportunity for additional insights,
as they allow us to study not only IP-to-IP traces but also rack-to-rack traffic.
Fig.~\ref{fig:takeaways} (b) shows the complexities 
of three different Facebook clusters, HAD, WEB, and DB.
First, we see that some applications (e.g., Hadoop) feature
much more structure than others (e.g., DB). However, 
we also observe the \emph{impact of aggregation}: 
at the rack level, we see a higher temporal complexity: communication
becomes more random. Moreover, WEB and DB have a slightly lower non-temporal complexity on the rack level,
an indication that this traffic has a high structure and placement in datacenters is subject to optimization.
In case of HAD, the rack-level complexity is higher than the 
IP-level complexity, in both dimensions: as we move higher to the topology, communication becomes more uniform.
It is important to note that the Facebook traces are sampled from real traffic, but only from a part of the racks in the datacenter~\cite{roy2015inside}. As a result, this trace is not a perfect representation of the entire network, especially in terms of its source distribution. To get more accurate results, without introducing under-estimation in non-temporal complexity, we modified the uniform transformation $\uni(\sigma)$ such that it will generate the source and destination columns individually, only from the set of IDs found in the its respective columns.
Also, we note that while sampling influences the measured absolute values, 
it does not affect the \emph{relative} conclusions, e.g., regarding the relatively higher complexity observed on the rack-level.

\para{The complexity of sources and destinations is unique to each trace.}
So far, we have focused on communication \emph{pairs}, however, our methodology
also allows us to investigate the complexities introduced by sources 
and destinations separately. Indeed, traffic matrices may be asymmetric
in that while communication sources are more skewed in the traffic trace,
communication destinations are uniform, and vice versa.
Accordingly, Fig.~\ref{fig:takeaways} (c) depicts the complexity map for different traces separated by their source and destinations.
In case of pFabric traces, sources and destinations behave similarly, 
which is expected, given that they are sampled uniformly at random. 
In the case of HPC, the non-temporal complexity is high, namely, the work seems to be uniformly divided among all CPUs. See also the marginal source and destination distributions in 
Fig.~\ref{fig:HPCmat}(a).
Interestingly, the sources and destinations individually contain    
temporal structure (where the source has lower complexity), which may be an indication that operations proceed in rounds, e.g., where a node (CPU) first sends several requests and then receives answers asynchronously.
The IP traces for WEB application reveal that the sources have less temporal complexity
than the destinations, which may be explained by the star-like communication patterns of
a web server; the lower non-temporal complexity indicates a more skewed popularity distribution
of web servers (compared to cache destinations which are load-balanced). 
The high non-temporal complexity of Hadoop on the rack-level shows the 
rather equal distribution, however, temporal structure may be leveraged due to 
consecutive communications. The low non-temporal complexity of destinations
is a result of the \emph{outband} sampling of the FB trace where the number of sources is much smaller than the number of destinations. 

\section{Traffic Generation Model}
\label{sec:model}

One interesting implication of our methodology is that
it naturally lends itself as the basis for synthetic
traffic generators. Furthermore, it allows us to provide
formal guarantees on the complexity of stochastic processes. In the following, we discuss these two
aspects.

\subsection{From Analysis to Synthesis}

We next tackle the question of how to \emph{synthesize}
traffic workloads of a particular temporal and non-temporal complexity.
Given the limited amount of publicly available communication traces, 
such a model can be particularly useful to generate
synthetic benchmarks, allowing researchers to compare
their algorithms in different settings (e.g., for
longer communication traces). Hence, in the following,
we propose an approach which allows to efficiently generate traces with
formal guarantees on their expected complexity for any
specific point on the complexity map. It is important to note that for all points on 
the map, there could be many traces (and models) 
whose complexity maps to this point. Our model provides one such solution.

To derive formal guarantees, we propose a simple Markovian model which is a stationary random process with a well-defined entropy rate~\cite{cover2012elements}. 
The model has two components: temporal and non-temporal. 
The non-temporal component is a joint probability traffic matrix, $M$, which can be computed from a given trace $\sigma$, similar to Fig.~\ref{fig:intro2}(c), or represent a known distribution (e.g. Zipf) where the entropy depends on the distribution's parameters. 
The temporal component is a repeating probability $p$.
To generate a trace, we start by sampling the first pair from $M$, and then at each step we add a pair to the trace, with probability $p$ we repeat the last pair and with probability $1-p$ we sample a new pair from $M$. 
More formally, to emulate a point $(x,y)$ on the complexity map we set $H(M) = y \cdot 2\log n$ where $H(M)$ is the joint entropy of $M$. It can be shown that $M$ is the stationary distribution of the chain and the non-temporal complexity of the model is $$y=H(M)/(2\log n)$$ The temporal complexity of the model is 
$$x=\frac{H(p,1-p) + (1-p)H(M)}{H(M)}$$ 
\noindent where $p$ can be computed analytically given $x$. Therefore, we can produce traces with similar complexities on the complexity map.

 \begin{figure}[t]
   \begin{centering}
   \includegraphics[width=.6\columnwidth]{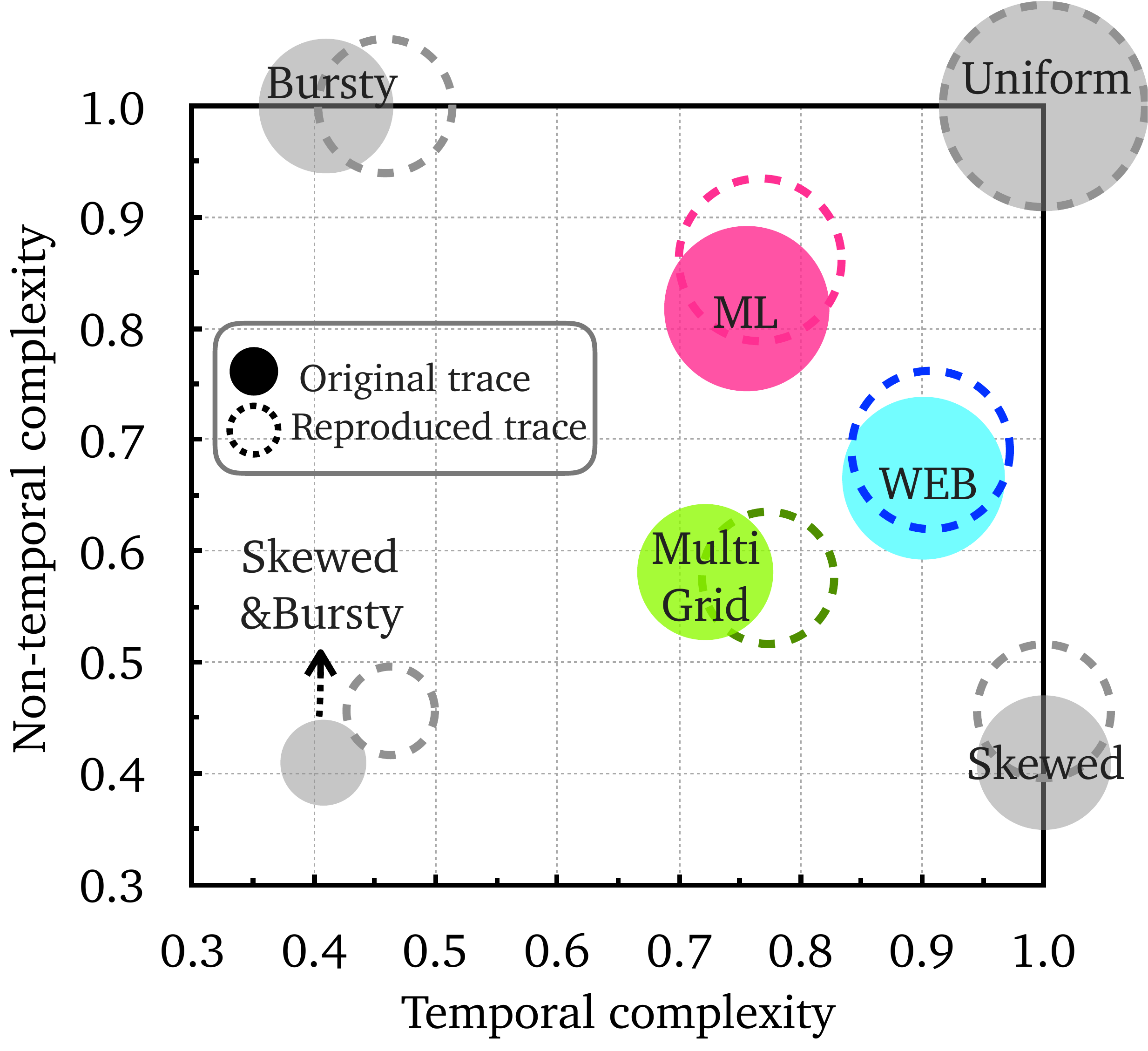} 
     \caption{Our traffic generation model is able to a produce new trace files with similar complexities as the original traces. The solid circles indicate the original trace's complexity and dashed circles represent the complexity of the trace produced by our traffic generator.}
     \label{fig:model}
   \end{centering}
 \end{figure}

Figure~\ref{fig:model} presents the quality of the above traffic generation model in producing syntactic traces for seven example points in the complexity map. First we used the model to reproduce the four hypothetical points from Fig.~\ref{fig:example_complexity_map}: \emph{Uniform}, \emph{Skewed}, \emph{Bursty}, and \emph{Skewed \& Bursty}. 
As a skewed distribution we used Zipf distribution with an exponent parameter of $\frac{2}{3}$ which leads to a normalized entropy of $0.4$ for a trace that is based on 16 IDs and 256 unique source-destination pairs.
For the other three traces: ML, MultiGrid, and WEB, we used  the joint probability  matrix of each trace.  By applying the compression methodology to the synthetic traces generated using our model, we find that our empirical traces are relatively close to the original trace files in the complexity map. However, the quality of these results is practically bounded by the selection of the compressing method and the ability of the compressor to capture precisely the entropy rate. Moreover, currently we do not reproduce exact packet arrival times, rather we take the order of packets as a proxy for temporal structure. In future work, we plan to add packet arrival times and flow-level information to the complexity analysis and our traffic generation model.

\vspace{-0.2cm}
\subsection{Supporting Formal Analysis}
\label{sec:formal_guarantee}

Our methodology can also provide a framework for formal analysis. 
In the following, assume that $\sigma$ is a trace generated by a 
\emph{stationary stochastic process}~\cite{cover2012elements}. Then, 
for long sequences and using
an optimal compression algorithm (such as the Lampel-Ziv \cite{ziv1977universal}),
we will achieve the compression limit defined by Shannon~\cite{cover2012elements, wyner1994sliding, vegetabile2017estimating}:  
the entropy rate of the process. 
From this, the normalized complexities of $\sigma$ can be proved analytically.

Let $\mathcal{Z} = \{Z_i\}$ be a stationary stochastic process 
that generates $\sigma$ where $Z_i=\{S_i, D_i\}$ are time-indexed random variables,
and $S_i \in S$ and $D_i \in D$ are a random source and a random destination at time $i$, respectively. Since $\mathcal{Z}$ is stationary, let $\pi$ denote the  stationary distribution and note that $\pi$ is a joint distribution over $S$ and $D$. With a slight abuse of notation, let $S$ and $D$ also denote the random variable of $\pi$. Note that $S$ and $D$ may be defined over different domains (IDs) and may be dependent. 
Also $Z_i$ may be dependent on a past element in $\mathcal{Z}$. 

A basic measure for the complexity of $\mathcal{Z}$ is based 
on Shannon Entropy and known as the \emph{entropy rate} \cite{cover2012elements},  $H(\mathcal{Z})$. The entropy rate of a stochastic process captures 
the expected number of bits per symbol (i.e., source or destination IDs)
that are both necessary and sufficient  to describe $\sigma$; or, alternatively, 
the expected amount of uncertainty in the next symbol given past symbols in the sequence.
The smaller the entropy rate, 
the less complex is the sequence (it requires fewer bits to describe/compress it). 

We can use the previous definitions of normalized trace complexity of $\sigma$ 
and formally relate them to entropy rates when $\sigma$ is generated by a stationary stochastic process.

\begin{theorem}[Trace Complexity Ratios of a Stationary Process] 
Consider an indexed stationary stochastic trace process $\mathcal{Z} = \{Z_t\}$ to generate $\sigma$ where $Z_t=\{S_t, D_t\}$ and $n=\lvert S \cup D \rvert$. If an optimal compression algorithm $C$ is used, then:
\begin{enumerate}
\item The \emph{trace complexity ratio} is
\begin{align}
\lim_{\substack{t \rightarrow \infty}} \comp(\sigma) = \frac{C(\sigma)}{C(\uni(\sigma))} =\frac{H(\mathcal{Z})}{2 \log n}
\end{align}

\item The \emph{temporal complexity ratio} is:
\begin{align}
\lim_{\substack{t \rightarrow \infty}} T(\sigma) = \frac{C(\sigma)}{C(\row(\sigma))} =\frac{H(\mathcal{Z})}{H(S,D)}
\end{align}

\item The \emph{non-temporal complexity ratio} is:
\begin{align}
\lim_{\substack{t \rightarrow \infty}} NT(\sigma) = \frac{C(\row(\sigma)))}{C(\uni(\sigma))} = \frac{H(S,D)}{2 \log n}
\end{align}
\end{enumerate}
\end{theorem}

\section{Related Work}\label{sec:related}

Information-theoretic approaches
and compression methodologies have already been proven successful
in capturing entropy in other domains such as
email~\cite{bratko2006spam}, or comment~\cite{kantchelian2012robust} 
spam filtering, or estimating neural discharges~\cite{amigo2004estimating}.
The study of traffic patterns
and the design of models is an evergreen topic 
of high relevance in the networking literature,
and 
examples  where measurement studies
spurred much research into traffic modeling
dates back to 
the 1990s~\cite{likhanov1995analysis,crovella1997self,leland1994self}. 
Since then, a large number of
methodologies have been developed~\cite{park1996relationship,taqqu1995estimators,shriver1998analytic},
based on temporal statistics~\cite{arima,wavelet,bmodel},
spatial statistics~\cite{cressie1992statistics,cox1980point},
and physical~\cite{barford1998generating} and 
information-theoretic~\cite{towsley1,towsley2} models.  
In contrast to prior work,
we are primarily interested in
the communication pattern itself,
rather than in the volume or headers of the exchanged data.
While our work builds upon many significant results
developed over the last decades~\cite{ziv1978compression}, 
we are not aware of any work which allows to systematically
differentiate between temporal and non-temporal components of
traffic traces. 

\section{Conclusion}\label{sec:conclusion}

The specific characteristics of traffic workloads have important implications for emerging network fabrics and algorithms~\cite{roy2015inside}.
This paper takes the first steps at understanding the complexity of traffic traces. In addition to temporal and non-temporal complexity measures, our entropy-based approach can be used to investigate other dimensions of the trace structure, e.g., regarding source-destination dependencies,
or to explore structure in transmission times and packet headers. 
More generally, while trace structure indicates potential for 
optimizations, it remains to develop algorithms which
exploit this structure to improve network performance and/or utilization.
 
{
  \bibliographystyle{ieeetr} 
   \balance
\bibliography{literature}
}

\end{document}